\documentclass[journal]{IEEEtran} 

\usepackage{amsmath,amsfonts}
\usepackage{xcolor}
\usepackage[ruled,vlined]{algorithm2e}
\usepackage{algpseudocode}
\usepackage{array}
\usepackage{flushend}

\usepackage{textcomp}
\usepackage{stfloats}
\usepackage{url}
\usepackage{verbatim}
\usepackage{graphicx}
\usepackage{cite}
\usepackage{subfigure}
\usepackage{caption}
\usepackage{threeparttable}
\usepackage{booktabs}
\usepackage{multirow}
\usepackage{hyperref}
\usepackage{color}

\newcommand{\eg}{e.\,g.,\ }
\newcommand{\ie}{i.\,e.,\ }

\begin{document}

\title{Re-Parameterization of Lightweight Transformer for On-Device Speech Emotion Recognition}

\author{Zixing~Zhang,~\IEEEmembership{Senior Member,~IEEE,}
        Zhongren~Dong,~\IEEEmembership{Student Member,~IEEE,}
        Weixiang~Xu, \\
        Jing~Han,~\IEEEmembership{Senior Member,~IEEE}
\thanks{The work was supported by the Guangdong Basic and Applied Basic Research Foundation under Grant Nr.~2024A1515010112 and the Changsha Natural Science Foundation under Grant Nr.~kq2402082. (\textit{Corresponding author: Jing~Han})}
\thanks{Z.~Zhang, Z.~Dong, and W.~Xu are with the College of Computer Science and Electronic Engineering, Hunan University, Changsha, 410082, China; Z.~Zhang is also with the Shenzhen Research Institute, Hunan University, Shenzhen 518000, China. \{zixingzhang, zrdong, xuweixiang\}@hnu.edu.cn}
\thanks{J.~Han is with the Department of Computer Science and Technology, University of Cambridge, 15 JJ Thomson Ave, Cambridge CB3 0FD, UK. jh2298@cam.ac.uk}
}




\maketitle

\begin{abstract}
With the increasing implementation of machine learning models on edge or Internet-of-Things (IoT) devices, deploying advanced models on resource-constrained IoT devices remains challenging. Transformer models, a currently dominant neural architecture, have achieved great success in broad domains but their complexity hinders its deployment on IoT devices with limited computation capability and storage size. Although many model compression approaches have been explored, they often suffer from notorious performance degradation. 
To address this issue, we introduce a new method, namely Transformer Re-parameterization, to boost the performance of lightweight Transformer models. It consists of two processes: the High-Rank Factorization (HRF) process in the training stage and the de-High-Rank Factorization (deHRF) process in the inference stage. In the former process, we insert an additional linear layer before the Feed-Forward Network (FFN) of the lightweight Transformer. It is supposed that the inserted HRF layers can enhance the model learning capability. In the later process, the auxiliary HRF layer will be merged together with the following FFN layer into one linear layer and thus recover the original structure of the lightweight model. 
To examine the effectiveness of the proposed method, we evaluate it on three widely used Transformer variants, \ie ConvTransformer, Conformer, and SpeechFormer networks, in the application of speech emotion recognition on the IEMOCAP, M$^{3}$ED and DAIC-WOZ datasets. Experimental results show that our proposed method consistently improves the performance of lightweight Transformers, even making them comparable to large models. The proposed re-parameterization approach enables advanced Transformer models to be deployed on resource-constrained IoT devices.
\end{abstract}


\section{Introduction}
\IEEEPARstart{W}{ith} the proliferation of Internet-of-Things (IoT) devices, deploying artificial intelligence (AI) models on IoT devices becomes imperative to enable real-time intelligent services while preserving privacy \cite{sharma_optimal, zhang_empowering, cai_enable, sun_tinyAD}. Speech emotion recognition (SER) is a typical application that needs to be deployed on IoT devices, such as smart home assistants and healthcare wearables, due to its privacy and real-time requirements \cite{chatterjee_real, tariq_speech, damilola_iot,shamim_emotion}.
After the inception of Transformer~\cite{vaswani_attention}, it has received tremendous success in natural language processing (NLP), speech processing, and computer vision, and continuously achieves state-of-the-art (SOTA) performance in diverse applications, such as emotion recognition, language translation, dialogue systems, and speech recognition~\cite{Qiu20-Pre,Khan22-Transformers,wagner_dawn}. Such great success is not only due to the considerable increase of training data and the optimization of Transformer architecture, but also largely attributed to the remarkable expansion of model size~\cite{Qiu20-Pre,Zhao23-Survey}. For example, the Transformer decoder-based language model of GPT-3 has 175 billion parameters~\cite{Brown20-language} and the model Pangu-$\Sigma$ has even more than one Trillion parameters~\cite{Ren23-Pangu}. The contribution of model size to the model performance has been comprehensively and empirically investigated in NLP~\cite{Hoffmann22-Training}.

Meanwhile, in many scenarios, it is needed to deploy models on IoT devices because of the privacy and security, and low-latency requirements~\cite{Deng20-Model,Li20-Federated,maruf_efficient,zhang_toward}. For example, the tasks of emotion recognition and health screening and diagnosis, just to name a few, highly relate to users' personal information, and require keeping the data on their own devices to protect their privacy and security. Besides, other tasks such as command recognition, face detection, and autonomous driving system are always on and require real-time decision-making. However, these IoT devices often lack sufficient storage memory, computational resources, energy, and network bandwidth. 

The contradiction between the complexity (\eg model size and computational operations) of Transformer models and the resource-constrained IoT devices has crucially hindered the on-device deployment of Transformer-style models. To deal with this challenge, plenty of model compression and optimization approaches have been introduced to date~\cite{ganesh_compressing,tay_efficient}, including pruning \cite{lecun_optimal}, knowledge distillation \cite{hinton_distilling}, and matrix decomposition~\cite{wang_linformer}. However, with these model compression approaches, especially when compacting the model into a tiny size one, the performance of the lightweight Transformer  often significantly degrades due to the reduction of its learning capability~\cite{Deng20-Model,Zhang22-Wakeupnet}.

To address this issue, we propose a novel approach, namely \textit{Transformer re-parameterization}, to boost the performance of lightweight Transformer models without any increase of model size and computational operations. This Transformer re-parameterization approach comprises two processes, \ie the High-Rank Factorization (HRF) process in the training stage and the de-High-Rank Factorization (deHRF) process in the inference stage. Specifically, in the HRF process, we insert an expanded linear layer before a feedforward network (FFN) of the Transformer. By doing this, it can endow the model with a better learning capability in the training stage due to the increase of model parameters. 
This process can be algebraically reversed. Thus, in the inference stage, the inserted linear layer will be removed by merging it together with its followed dense layer into one new dense layer with a mathematical calculation. By this means, the original lightweight Transformer architecture will be reconstructed without adding new parameters and computational operations in the inference stage whilst maintaining the boosted model performance. We evaluate the introduced approach in SER due to its broad applications, such as human-machine interaction, call centers, and mental health tracking~\cite{Han17-Hard,wagner_dawn}. 

The present work is highly motivated by the `scaling law' of deep models~\cite{Kaplan20_SCaling,Henighan20_Scaling}, where the model performance has a power-law relationship with the model size. This is not only empirically demonstrated via various tasks and models~\cite{Kaplan20_SCaling,Henighan20_Scaling}, but also mathematically explained through the theory that an interpolated/over-parameterized classifier is more generalized to the unseen data than a small model~\cite{Ma18_The}. 
In this work, we focus on the Transformer models and investigate their different modules, including the queries, keys, and values (QKV) for attention mechanisms, Projection, FFN, classification head (CLS), and their combinations. To the best of our knowledge, this is the first work for enhancing the lightweight Transformers via the lossless model re-parameterization strategy. 

Our major contributions  can be summarized as follows:
\begin{itemize}
    \item We, for the first time, introduced a novel re-parameter-ization strategy for enhancing the lightweight Transformers, to the best of our knowledge. It empowers lightweight Transformers with better learning capability in the training process whilst no performance loss in the inference stage, without the price of model size and computational operation. 
    \item We comprehensively investigated the proposed re-parameterization strategy in different modules of Transformer, including QKV for attention mechanisms, projection, FFN, CLS, and their combinations. 
    \item We extensively evaluate the effectiveness and generalizability of our approach in the context of SER, a domain with a strong imperative for on-device deployment due to privacy concerns.
\end{itemize}

The remainder of this paper is organized as follows. We first introduce related work in Section~\ref{section:2}, and then formalize the problem statement and elaborately describe the proposed method in Section \ref{section:3}. After that, the experimental setups and experimental results are given in Section \ref{section:4}. Finally, we draw the conclusion in Section \ref{section:5}.

\section{RELATED WORK}\label{section:2}
In this section, we review Transformer-based SER, lightweight Transformers, and structural re-parameterization.

\subsection {Transformers-based Speech Emotion Recognition}
For SER, nowadays Transformer-based models have been widely employed due to their capability to capture short- and long-context information. From the feature aspect, it goes from the handcrafted frame-level features (\eg the frame-level Mel-Frequency Cepstral Coefficient [MFCC] or Log Mel-filter Bank energies [LMFB])~\cite{Nediyanchath_Multi} and the segment-level features (\eg IS09 or eGeMAPS)~\cite{Schuller13-INTERSPEECH}, to the raw speech signals that contain the complete information in an end-to-end way~\cite{tarantino_self}. The latter, however, often require a large number of emotional speech data for training.

From the model architecture aspect, early studies often used the classic Transformer for SER~\cite{Nediyanchath_Multi,tarantino_self}. Recently, more and more advanced Transformer structures were designed for SER. For example, a 1D CNN network was inserted into the classic Transformer to extract information from raw speech signal~\cite{tarantino_self}. Besides, a SpeechFormer model was introduced, which optimized global attention to compute local attention and employed a hierarchical structure to extract frame-, phoneme-, word-, and sentence-level features, resulting in superior performance~\cite{chen_speechformer}. Furthermore, a DWFormer model has been proposed, which takes dynamic windows to extract fine-grained temporal features from speech samples for SER~\cite{chen_dwformer}.

Notably, all these aforementioned studies trained Transformer models from scratch in a supervised learning way. With the advent of pre-trained models, nowadays more and more research is focusing on the use of pre-trained models for SER~\cite{wang_fine,wagner_dawn}. For example, Wang et al.~\cite{wang_fine} performed partial and overall fine-tuning of pre-trained models, including Wav2vec 2.0 and HuBERT, to study their feasibility in SER, speaker verification, and spoken language understanding tasks. 
Wagner et al.~\cite{wagner_dawn} conducted a comprehensive evaluation of several pre-trained variants of Wav2vec 2.0 and HuBERT for SER. Although these methods show promising performance, their high computational complexity, and particularly their model size, renders them unsuitable for edge devices.

\subsection {Lightweight Transformers}
With the popularity of mobile/edge devices, increasing interest is focusing on the reduction of the computational cost and model size of Transformers and have proposed many compression methods \cite{michel_sixteen, voita_analyzing, prasanna_when, chen_earlybert, fan_reducing, sajjad_effect, sanh_distilbert, tinyBERT, mobileBERT, distilhubert, wang_linformer, tan_fmmformer, geng_is, h_drone, winata_lightweight, ren2022t}. These methods can be mainly classified into three categories: pruning \cite{michel_sixteen, voita_analyzing, prasanna_when, chen_earlybert, fan_reducing, sajjad_effect}, knowledge distillation \cite{sanh_distilbert, tinyBERT, mobileBERT, distilhubert} and matrix decomposition \cite{wang_linformer, tan_fmmformer, geng_is, h_drone, winata_lightweight, ren2022t}.

\begin{figure*}[!t]
    \centering
    \subfigure[Pruning]{
    \label{fig:pruning}
        \includegraphics[width=0.23\textwidth]{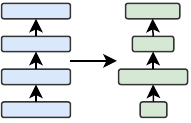}}
    \subfigure[Knowledge Distillation]{
    \label{fig:knowledge_distillation}
        \includegraphics[width=0.23\textwidth]{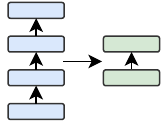}}
    \subfigure[Matrix Decomposition]{
    \label{fig:matrix_decomposition}
        \includegraphics[width=0.235\textwidth]{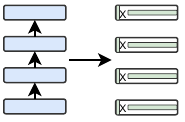}}    
    \subfigure[Structural Re-Parameterization]{
    \label{fig:re_parameterization}
        \includegraphics[width=0.25\textwidth]{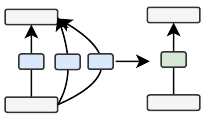}}  
    \caption{Four categories of model compression methods. (a) Pruning: Removing some weights or connections from the network.
    (b) Knowledge Distillation: Transfering knowledge from a large and accurate teacher model to a smaller student model.
    (c) Matrix Decomposition: Decomposing a large weight matrix into several smaller matrices.
    (d) Structural Re-Parameterization: Shrinking a multi-branch structure into a single-branch structure, typically employed in convolutional neural networks.
    }
    \label{fig:lightweight_transformer}
\end{figure*}

Pruning removes redundant connections from the model, and thus reduces the model size and computation operations, and speeds up the inference process. The simplified process of pruning refers to Figure~\ref{fig:pruning}. Studies on Transformer-based pruning typically concentrate on different modules of Transformer, such as pruning the attention head \cite{michel_sixteen, voita_analyzing}, pruning the FFN \cite{prasanna_when, chen_earlybert}, or even directly removing certain layers \cite{fan_reducing, sajjad_effect}. Researchers often carefully eliminate a module or layer based on its contribution to the model size, ensuring that the overall performance of the model is not compromised by excessive removal.

Knowledge distillation trains a smaller and less computationally intensive model by transferring knowledge from a large and more accurate model. Its streamlined procedure is illustrated in Figure~\ref{fig:knowledge_distillation}. One of the most basic purposes of Transformer-based distillation work is to integrate the capabilities of the teacher model into a relatively smaller model. DistillBERT~\cite{sanh_distilbert} and TinyBERT~\cite{tinyBERT} both integrate standard BERT into a model with fewer layers, and MobileBERT~\cite{mobileBERT} integrates standard BERT into a narrower network, with the same number of layers as BERT. Similarly, Distilhubert~\cite{distilhubert} has the standard structure of HuBERT but fewer layers, retaining almost the same performance as HuBERT. One feature of knowledge distillation is that no matter how small your student model is, we need to train a large teacher model.

Matrix decomposition aims to reduce the number of parameters and computation by decomposing the weight matrix in the model into the product of multiple successive sub-matrices, and its simplified process refers to Figure~\ref{fig:matrix_decomposition}. Early work on transformer-based matrix decomposition typically focused on approximating the attention matrix using a low-rank matrix for the self-attention module~\cite{wang_linformer, tan_fmmformer, geng_is}. Subsequently, researchers went beyond the attention module and performed matrix decomposition on the entire transformer to compress the model, potentially achieving greater compression~\cite{h_drone, winata_lightweight, ren2022t}. Matrix decomposition does not decrease the number of layers in the model but only approximates certain modules in the transformer, such as the self-attention module and the FFN module. Consequently, matrix decomposition methods usually do not compress the model to a significant extent. As the degree of compression increases, the model's performance tends to decrease substantially.

In any scenario, despite the reduction in the size of the network through the aforementioned compression methods, it not only fails to provide the flexibility to design a network with a specific architecture but also does not combine excessive parameterization to enhance the performance of training a compact network \cite{guo_expandnets}.

\subsection {Structural Re-Parameterization}
Structural re-parameterization \cite{ding_repvgg} is different from all aforementioned methods. 
It decouples the model training and inference stages, typically employing a multi-branch structure during the training process, and effectively reverts back to the original single-branch structure during the inference phase. The simplified process of structural re-parameterization refers to Figure~\ref{fig:re_parameterization}. Structural re-parameterization was previously investigated in CNN networks, such as using asymmetric convolution instead of regular convolutional layers \cite{ding_acnet}, expanding a single regular convolution into multiple convolutional layers \cite{guo_expandnets}, and expanding a regular convolution into a multi-branch convolution \cite{ding_repvgg}. A common feature of these networks is that the expanded modules can be shrunk back into a single-branch standard convolution in the inference stage, which allows the model to be characterized by both representational power and computational efficiency. To date, no structural re-parameterization techniques have been studied for boosting the lightweight Transformer performance.

\begin{figure*}[!t]
    \centering
    \subfigure[ConvTransformer/SpeechFormer]{
    \label{fig:hrf_conv_speechformer}
        \includegraphics[width=0.5\textwidth]{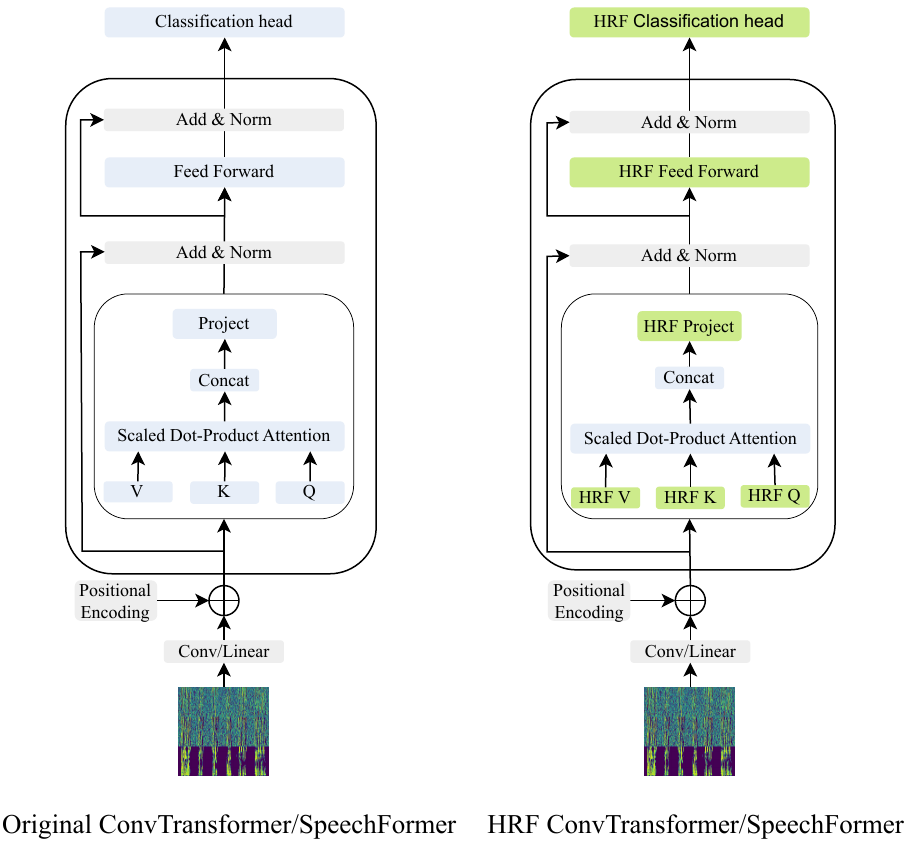}}
    \subfigure[Conformer]{
    \label{fig:hrf_conformer}
        \includegraphics[width=0.4\textwidth]{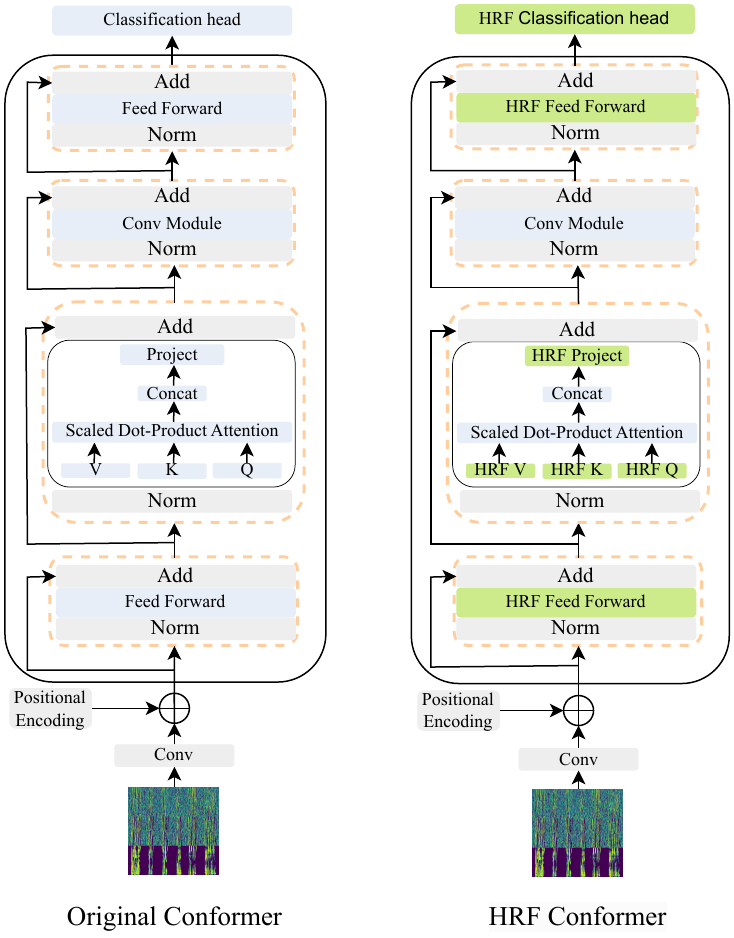}}
    \caption{Framework of Transformer re-parameterisation in different modules with the examples of ConvTransformer, SpeechFormer, and Conformer.}
    \label{fig:overview}
\end{figure*}

\section{RE-PARAMETERIZATION OF LIGHTWEIGHT TRANSFORMER}\label{section:3}

In this section, we first formalize the problem statement, then present the framework overview, and finally describe the detailed process and implementation rationale for the Transformer re-parameterization.

\subsection{Problem Formulation}
We evaluate the proposed Transformer re-parameterization in the application of SER. 
Let $D=\left \{ \left ( x_{1},y_{1} \right ),\left ( x_{2},y_{2} \right ),\cdots , \left ( x_{n},y_{n} \right ) \right \} $ be the dataset consisting of $n$ utterances, where $x_{i}$ denotes the $i^{th}$ utterance and $y_{i}$ denotes its corresponding emotion label (\eg neutral, sadness, happiness, fear, anger, surprise, and disgust). The goal is to train a model $\mathcal{M}$ on $D$ to predict the emotion labels of the test utterances.

\subsection{Overview}\label{subsec:overview}
The overall framework diagram is shown in Figure~\ref{fig:overview}, where we select three widely used Transformer structures in the speech domain, \ie ConvTransformer~\cite{yeh_transformer}, Conformer~\cite{gulati_conformer}, and SpeechFormer~\cite{chen_speechformer}. The introduced Transformer re-parameterization includes High-Rank Factorization (HRF) process in the training stage and de-High-Rank Factorization (deHRF) process in the inference stage.

Specifically, in the training stage: For ConvTransformer or SpeechFormer, we perform HRF on QKV module, Project module, FFN module, CLS module, and all modules (ALL). As the FFN module is composed of two fully connected (FC) layers, HRF FFN will have three strategies, \ie FFN-1, FFN-2, and FFN-1 \& FFN-2. Detailed information is illustrated in the right subfigure of Figure~\ref{fig:hrf_conv_speechformer}. 
Similar to ConvTransformer and SpeechFomer, we apply HRF on the same modules for Conformer. 
Nevertheless, as a Macaron structure is performed for Conformer, \ie two FFN networks are used before and after the self-attention block, we additionally conduct HRF on the additional FFN module (FFN\_M) as well, \ie FFN\_M-1, FFN\_M-2, and FFN\_M-1 \& FFN\_M-2. More details about the HRF Conformer can be found in the right subfigure of Figure~\ref{fig:hrf_conformer}.

In the inference stage: All the HRF modules will be reconstructed and converted back to the original structure, which is presented in the left subfigures of Figure~\ref{fig:hrf_conv_speechformer} for ConvTransformer or SpeechFormer and Figure~\ref{fig:hrf_conformer} for Conformer. 

\begin{figure}[!t]
  \centering
  \includegraphics[width=\linewidth]{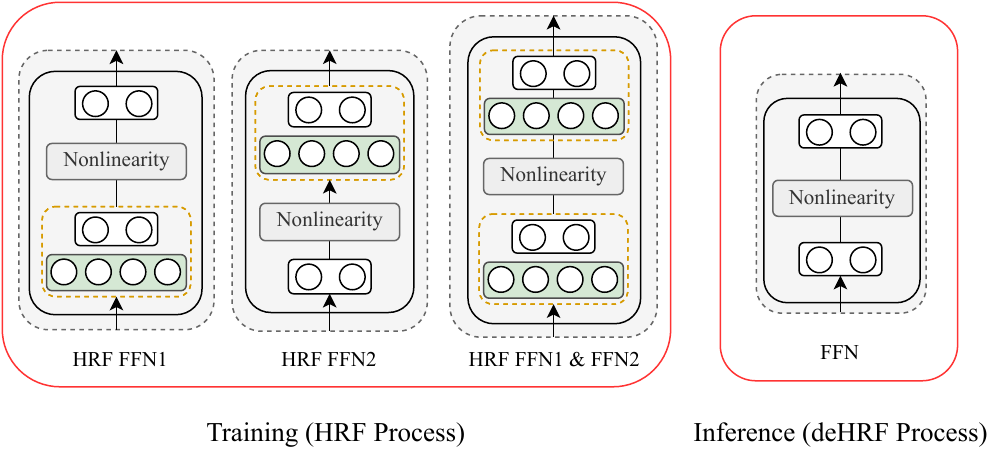}
  \caption{Detailed illustration of the re-parameterization process in the Transformer Feed-Forward Network (FFN).}
  \label{HRF_process}
\end{figure}

Taking an FFN module of Transformer for example, Figure~\ref{HRF_process} depicts detailed HRF and deHRF processes.  In the training phase, three types of HRF expansions are performed, \ie FC1 only (FFN-1), FC2 only (FFN-2), and  both FC1 and FC2 (FFN-1 \& FFN-2). In the inference phase, the expanded FC layers are shrunk to a single FC layer by a deHRF process. The principles of HRF and deHRF are explained in Section \ref{HRF} and Section \ref{deHRF}.

\subsection{Linear Over-Parameterization} \label{Linear}

Before introducing our method, let's first qualitatively discuss the importance of linear overparameterization, which is also the theoretical basis of HRF. Suppose we are learning a linear model parameterized by a matrix $W$, obtained through training $\min \left \{ L\left (  W\right )  \right \}$. Now, we use a linear neural network with N layers to implement the learning process, then $W=W_{N}W_{N-1}\dots W_{1}$, where $W_{j}$ is the weight matrix of a specific layer. We use a lower learning rate $\eta$ for gradient descent, and according to \cite{arora2018optimization}, the weight matrix $W=W_{N}W_{N-1}\dots W_{1}$ satisfies the dynamics of continuous gradient descent:

\begin{equation}
  W_{T}^{j+1} W_{j+1}=W_{j}W_{j}^{T}.
\end{equation}

Then the update rule for the end-to-end weight matrix can be expressed in the form of a differential equation:

\begin{equation}
  W^{t+1}\longleftarrow W^{t}-\eta \sum_{j=1}^{N}\left [ W^{t}\left ( W^{t} \right )^{T}   \right ]^{\frac{j-1}{N}}\frac{\mathrm{d} L}{\mathrm{d} W}\left ( W^{t} \right )  \left [ \left ( W^{t} \right )^{T}W^{t}   \right ]^{\frac{N-j}{N}},
\end{equation}
where $t$ is now a discrete time index, $\left [ \cdot \right ]^{\frac{j-1}{N}}$ and $\left [ \cdot \right ]^{\frac{N-j}{N}}$,$j=1,\cdots N$ are fractional power operators defined on positive semi-definite matrices. We left multiply by a matrix $\left [ W^{t}\left ( W^{t} \right )^{T}   \right ]$, right multiply by a matrix $\left [ \left ( W^{t} \right )^{T}W^{t}   \right ]$, and then sum over $j$, which is essentially a special form of preconditioning that promotes movement along the optimization direction and can be used for optimizing deep networks. More importantly, the above update rule does not depend on the width of the hidden layers in the linear neural network, but on the depth ($N$). This can be explained as overparameterization promoting movement along the direction already taken for optimization and can thus be viewed as a form of acceleration \cite{arora2018optimization}.

In addition to the qualitative discussion of overparameterization, in practice, the performance of large language models strongly depends on the number of model parameters N (excluding embeddings), the size of the dataset D, and the computational resources C used for training \cite{henighan2020scaling}. As long as we simultaneously increase N and D, the performance will predictably improve, which is known as the scale law of large language models \cite{kaplan2020scaling,hernandez2021scaling}. The scale law in practice demonstrates the importance of overparameterization for model performance.

\subsection{High-Rank Factorization} \label{HRF}
Theoretically, with the HRF processing, a FC layer that serves as the foundational layer in Transformer can be expanded into multiple FC layers. Assuming a FC layer with $m$ input neurons and $n$ output neurons, the output $Y$ can be represented as:
\begin{equation}
  Y=\sigma(WX+b), 
\end{equation}
where $X$ is the input, $Y$ is the output of the FC layer, $\sigma()$ is the activation function, $W\in R^{m\times n}$ is the weight matrix, and $b\in R^{n}$ is the bias. The HRF process is defined as follows:
\begin{equation}
  W= W_{N}\times W_{N-1}\times\dots W_{1}.
\end{equation}
$N$ can be arbitrarily large. But, due to the computational cost of FC layers, we set the maximum $N$ to 3 in this paper. When $N=3$, $W_{3}\in R^{m\times r}$, $W_{2}\in R^{r\times r}$, $W_{1}\in R^{r\times n}$,  $r$ is the expansion scale and can be set to 2, 4, or 8 times the second dimension of $W$, \ie $r=2n$, 4n, or 8n. Notably and importantly, when expanding one FC layer into multiple FC layers, no activation functions are applied after each extended HRF layer. 

This process is similar to the low-rank matrix factorization/de-composition (LRF) that is widely used for model compression~\cite{Deng20-Model}: given any matrix $W^{*}\in R^{m\times n}$ of full rank $r^{*}\le\min\{m,n\} $, it can be decomposed into $W^{*}\simeq AH$, where $A\in R^{m\times r^{*}}$ and $H\in R^{r^{*}\times n}$. When $r^{*}$ is much smaller than $m$ or $n$, the space complexity can be significantly reduced from $O(mn)$ to $O(r^{*}(m + n))$. Because both LRF and HRF use multiple matrices to represent a single matrix with the difference that LRF has $r^{*}\le \min \{m,n\}$ whilst HRF normally has $r > \max \{m,n\}$, we refer to our proposed method as HRF. The details of HRF are shown in Algorithm~\ref{alg:hrf}.


\begin{algorithm}[!t]
\caption{Pseudocode of the HRF}\label{alg:hrf}
\begin{algorithmic}
\State \textbf{Input}:  \\
\hspace*{\algorithmicindent} $m$: Input dimension of FC layer \\
\hspace*{\algorithmicindent} $n$: Output dimension of FC layer \\
\hspace*{\algorithmicindent} $r$: Expansion ratio of HRF layer 
\State \textbf{Output}: \\
\hspace*{\algorithmicindent} $HRF$: High Rank Factorization layer
\end{algorithmic}

\begin{algorithmic}[1]
\Procedure{HRF}{$m,n,r$}
    \State $FC \gets Linear(m, n, bias=True)$
    \State $HRF1 \gets Linear(m, r \times n, bias=True)$
    \State $HRF2 \gets Linear(r \times n, n, bias=True)$
    \State $HRF \gets Sequential(HRF1, HRF2)$
    \State \textbf{return} $HRF$
\EndProcedure
\end{algorithmic}
\end{algorithm}

\subsection{De-High-Rank Factorization} \label{deHRF}
In the training phase, we employ HRF in the model; whilst in the inference phase, we reconstruct the HRF model back to its original size and structure. The reconstruction process involves a chain of matrix calculation as no non-linear activation functions are conducted between the adjacent FC layers. Given one FC layer with one HRF layer expansion (\ie $N=1$) for example: 
\begin{equation}
  Y=\sigma \left ( WX+b \right ),
\end{equation}
where $X$ is the input and $Y$ is the output. After applying HRF to FC layer, we obtain $Y^{*}$ that is calculated using the following equation:
\begin{equation}
  \begin{aligned}
        Y^{*}&=\sigma \left ( W_{2}\left ( W_{1}X+b_{1} \right ) + b_{2} \right )\\
          &= \sigma\left (\left(W_{2}W_{1}\right)X+\left ( W_{2}b_{1}+b_{2} \right )  \right ). 
    \end{aligned}
\end{equation}
Let $W=W_{2}W_{1}$, $b=W_{2}b_{1}+b_{2}$, then $Y=Y^{*}$. By doing this calculation, we can contract the two layers back to one original FC layer. Algorithm~\ref{alg:dehrf} describes the details of the deHRF process. To be noted, when $N>1$, a similar chain of matrix calculation can be conducted to reconstruct the HRF modules. 


\begin{algorithm}[!t]
\caption{Pseudocode of the deHRF}\label{alg:dehrf}
\begin{algorithmic}
\State \textbf{Input}:  \\
\hspace*{\algorithmicindent} $FC$: Original FC layer \\
\hspace*{\algorithmicindent} $HRF$: High Rank Factorization layer 
\State \textbf{Output}: \\
\hspace*{\algorithmicindent} $FC$: Re-constructed FC layer
\end{algorithmic}

\begin{algorithmic}[1]
\Procedure{deHRF}{$FC, HRF$}
    \State $HRF1, HRF2 \gets HRF$
    \State $W1, b1 \gets HRF1.weight, HRF1.bias$
    \State $W2, b2 \gets HRF2.weight, HRF2.bias$
    \State $W \gets MatrixMultiplication(W2, W1)$
    \State $b \gets MatrixMultiplication(W2, b1) + b2$
    \State $FC.weight.data \gets W$
    \State $FC.bias.data \gets b$
    
    \State \textbf{return} $FC$
\EndProcedure
\end{algorithmic}
\end{algorithm}

\section{EXPERIMENTS AND RESULTS} \label{section:4}

In this section, we introduce the selected datasets, experiment implementation details, and selected Transformer models, and present the results followed by extensive discussions and ablation studies.

\subsection {Datasets}

We evaluated the introduced Transformer Re-parameterization on two multimodal emotional datasets -- IEMOCAP \cite{busso_iemocap}, $M^{3}$ED \cite{zhao_m3ed} and DAIC-WOZ \cite{gratch2014distress}. 
In this work, we mainly focus on the speech signals for emotion recognition.

\textit{IEMOCAP} is the most widely exploited dataset for SER. It comprises 12.46 hours of audio data, divided into five sessions with one male and one female speaker each. The conversations were segmented into utterances and annotated by at least three annotators using the following discrete categories: anger, happiness, sadness, neutrality, excitement, frustration, fear, surprise, and others. For this study, we focused on the first four categories, and merged ``excitement'' into ``happiness''. Following previous research \cite{guo_emotion,ren_LRGCN}, we used the first four sessions for training and the last session for testing. To reduce randomness, we utilized five random seeds for training and testing, and averaged the final results over the five tests. We took unweighted accuracy (UA), weighted accuracy (WA), and weighted average F1 (WF1) as performance metrics for model evaluation.

\textit{$M^{3}$ED} is a recently released and the first Chinese multimodal sentiment dialogue dataset, comprising 990 dyadic emotional dialogues from 56 different TV series, with a total of 9\,082 turns and 24\,449 utterances. The dataset was annotated with seven emotional categories (\ie happy, surprised, sad, disgusted, angry, fearful, and neutral) at the verbal level, including sound, visual, and textual patterns. We followed the default data partition strategy and used five random seeds for training, validation, and testing, with the final results averaged over the five tests. We took WF1 as the performance metric to evaluate models. Note that the metrics of UA and WA were not considered for $M^{3}$ED mainly due to the alignment with other SOTA approaches for performance comparison. The detailed emotion distribution for the IEMOCAP and M3ED datasets are listed in Table \ref{tab:distribution_iemocap_m3ed}.

The DAIC-WOZ dataset, derived from The AVEC2017 series depression detection task, comprised 189 segments of clinical interviews. The dataset was divided into 107 segments for training, 35 segments for validation, and 47 segments for testing. These interviews were conducted to aid in diagnosing conditions such as depression.

Each segment in the dataset included audio and video features, alongside transcripts of the interviews. Segment durations ranged from 7 to 33 minutes, with an average duration of 16 minutes. To ensure fair comparisons with other state-of-the-art works, we utilized the training and validation sets and reported macro average F1 (MF1) scores on the validation set.

For training convenience, we extracted individual patient voices from the provided transcripts and segmented them into 10 s clips. The training set consisted of 5084 clips, while the validation set contained 1931 clips. Detailed distribution of the dataset is presented in Table \ref{tab:daic}.

\begin{table}[!t]
  \centering
  \caption{Emotion Distribution of IEMOCAP and M$^{3}$ED.} 
  \label{tab:distribution_iemocap_m3ed}
  \begin{threeparttable}
      \begin{tabular}{ccccccccc}
        \toprule
       \multirow{2}{*}{Emotion}    &\multicolumn{3}{c}{IEMOCAP}   & \multicolumn{3}{c}{M$^{3}$ED}\\
        \cmidrule(lr){2-4} \cmidrule(lr){5-7} 
                    &Train   &Val  &Test   &Train   &Val   &Test    \\
        \midrule
        neutral     &1,324 &- &384         &7,130   &1,043  &1,855   \\
        happy       &1,194 &- &442          &1,626   &303  &358   \\
        sad         &839 &- &245            &2,734   &489  &734   \\
        anger       &933 &- &170            &3,816   &682  &736   \\
        surprise    &- &- &-               &696   &120  &235   \\
        disgust     &- &- &-               &1,145   &134  &218   \\
        fear        &- &- &-               &280   &50  &65   \\
        \midrule
        Total       &4,290 &- &1,241     &17,427   &2,821  &4,201   \\
      \bottomrule
    \end{tabular}
\end{threeparttable}
\end{table}

\begin{table}[!t]
  \centering
  \caption{Data Distribution of DAIC-WOZ.}
  \label{tab:daic}
  \begin{threeparttable}
      \begin{tabular}{cccc}
        \toprule
         Class  &Train   &Val  &Test   \\
        \midrule
        N-Dep.*     &87 &28  &38 \\
        Dep.*      &21 &7  &9   \\
        \midrule
        Total Number of participants &107  &35  &47    \\
        Total Number of clips      &5084 &1931  &-     \\
      \bottomrule
    \end{tabular}
    \begin{tablenotes}  
      \item[1] 
        *All participants were assigned into one of two classes, depressed (Dep.) or non-depressed (N-Dep.), based on the PHQ-8 scores.
    \end{tablenotes}
\end{threeparttable}
\end{table}

\subsection{Implementation Details}
Generally, there are three types of models: the original model (the original large one), the lightweight model (a lighter version of the original one), and the HRF model (the lightweight model with an HRF process). We implemented all models under the PyTorch framework.

The initial learning rate for the original and lightweight ConvTransformer and Conformer were the same with 1e-3; whilst for the original and lightweight Speechformer they were 1e-3 and 5e-4, respectively. The learning rate decays to half when the loss does not decrease. 
In addition, the initial learning rate for the HRF model was identical to the corresponding lightweight model. All models used the AdamW optimizer, with the decay rate set to 1e-6 and the epoch set to 120 (for IEMOCAP) or 100 (for M$^{3}$ED and DAIC-WOZ). The original ConvTransformer, Conformer, and SpeechFormer had the structures of [8, 80, 320], [4, 80, 320], and [8, 80, 64], where the three values in each bracket denote the number of Transformer blocks ($n_{layer}$), the dimensionality of the input data for each Transformer block ($d_{model}$), and the dimensionality of the first FC layer in FFN ($d_{ffn}$), respectively. These original models were then compressed into tiny ones for the on-device deployment purpose, with new structures of [1, 16, 4], [1, 16, 2], and [1, 16, 4] for lightweight ConvTransformer, Conformer, and SpeechFormer, respectively. All models hold four attention heads. Notably, all HRF models had the same parameters as their corresponding lightweight models during the inference stage. The default activation functions for ConvTransformer and SpeechFormer are ReLU, and for Conformer it is Swish. 

Regarding to the acoustic features, we followed previous work~\cite{hou_semantic} and extracted 78-dimensional features from speech signals using 26 filter banks with a window length of 25\,ms and a hop size of 10\,ms. These 78-dimensional features comprise 26 original LMFBs, and their first and second derivatives. Please note that we do not directly employ the pre-trained models for feature extraction, as these pre-trained models are too large to be deployed on the device, or it is too sensitive to send the users' private data to the cloud. Besides, we segmented the sentences into 5\,s and 4\,s for IEMOCAP and M$^3$ED datasets, respectively, and pursued a zero-filling strategy if the sentence was insufficiently long. 

\begin{table*}[!t]
  \centering
  \caption{Performance of lightweight \textit{ConvTransformer} when  HRF is applied to its different modules  on the IEMOCAP, M$^3$ED and DAIC-WOZ datasets.}
  \label{tab:hrf_convtransformer}
  \begin{threeparttable}
      \begin{tabular}{cccccccccccc}
        \toprule
       \multirow{2}{*}{which modules use HRF [$\times$8]}    &\multicolumn{5}{c}{IEMOCAP}   & \multicolumn{3}{c}{M$^{3}$ED} & \multicolumn{3}{c}{DAIC-WOZ}\\
        \cmidrule(lr){2-6} \cmidrule(lr){7-9}  \cmidrule(lr){10-12}
        &WA     &UA     &WF1    &Params &FLOPs   &WF1    &Params &FLOPs  &MF1    &Params &FLOPs\\
        \midrule
        Original     &.589 &.602 &.586 & 801K  &226M  &.399 &595K  &160M &.572 &801K  &454M \\
        Lightweight  &.552 &.567 &.550 &9K     &7M    &.372 &9K    &6M   &.530 &9K    &14M \\
        \midrule
        FFN1         &.562 &.578 &.560 &9K     &7M  &.386 &9K    &6M  &.573 &9K    &14M\\
        FFN2         &\textbf{.580} &\textbf{.589} &\textbf{.580} &9K     &7M  &.382 &9K    &6M  &.558 &9K    &14M\\
        FFN1 \& FFN2  &.566 &.582 &.565 &9K     &7M  &.381 &9K    &6M &\textbf{.580} &9K    &14M\\
        QKV          &.566 &.583 &.562 &9K     &7M &.380 &9K    &6M &.551 &9K    &14M\\
        Project      &.566 &.583 &.565 &9K     &7M  &.382 &9K    &6M &\textbf{.580} &9K    &14M\\
        CLS      &.572 &.582 &.570 &9K     &7M  &\textbf{.387} &9K    &6M &.551 &9K    &14M\\
        ALL          &.565 &.571 &.563 &9K     &7M &.383 &9K    &6M &.530 &9K    &14M\\
      \bottomrule
    \end{tabular}
\end{threeparttable}
\end{table*}

\begin{table*}[!t]
  \centering
  \caption{Performance of lightweight \textit{Conformer} when  HRF is applied to its different modules on the IEMOCAP, M$^3$ED and DAIC-WOZ datasets.}
  \label{tab:hrf_conformer}
  \begin{threeparttable}
      \begin{tabular}{cccccccccccc}
        \toprule
       \multirow{2}{*}{which modules use HRF [$\times$8]}    &\multicolumn{5}{c}{IEMOCAP}   & \multicolumn{3}{c}{M$^{3}$ED} &\multicolumn{3}{c}{DAIC-WOZ}\\
        \cmidrule(lr){2-6} \cmidrule(lr){7-9} \cmidrule(lr){10-12}
        &WA     &UA     &WF1    &Params &FLOPs   &WF1    &Params &FLOPs &MF1    &Params &FLOPs\\
        \midrule
        Original     &.572 &.572 &.571 & 809K  &227M  &.378 & 323K  &50M  &.633 & 814K  &458M\\
        Lightweight  &.524 &.558 &.520 &10K   &7M  &.368 & 11K  &6M  &.533 & 10K  &15M\\
        \midrule
        FFN1         &.545 &.556 &.543 &10K   &7M  &.376 & 11K  &6M &.595 & 10K  &15M\\
        FFN2         &.545 &.560 &.542 &10K   &7M  &.375 & 11K  &6M  &.571 & 10K  &15M\\
        FFN1 \& FFN2  &.545 &.555 &.542 &10K   &7M  &.371 & 11K  &6M &.592 & 10K  &15M\\
        FFN\_M1  &\textbf{.554} &.557 &\textbf{.553} &10K   &7M  &\textbf{.379} & 11K  &6M &.603 & 10K  &15M\\
        FFN\_M2  &.545 &.553 &.542 &10K   &7M  &.376 & 11K  &6M &.596 & 10K  &15M\\
        FFN\_M1 \& FFN\_M2  &.542 &\textbf{.561} &.540 &10K   &7M  &.372 & 11K  &6M &.583 & 10K  &15M\\
        FFN1 \& FFN2 \& FFN\_M1 \& FFN\_M2   &.551 &.555 &.550 &10K   &7M   &.375 & 11K  &6M&\textbf{.644} & 10K  &15M\\
        QKV          &.536 &.535 &.532 &10K   &7M &.370 & 11K  &6M &.608 & 10K  &15M\\
        Project      &.547 &.560 &.545 &10K   &7M  &.375 & 11K  &6M &.603 & 10K  &15M\\
        CLS      &.535 &.555 &.532 &10K   &7M  &.367 & 11K  &6M &.540 & 10K  &15M\\
        ALL          &.548 &.558 &.547 &10K   &7M  &.367 & 11K  &6M &.570 & 10K  &15M\\
      \bottomrule
    \end{tabular}
\end{threeparttable}
\end{table*}

\begin{table*}[!t]
   \centering
   \caption{Performance of lightweight \textit{SpeechFormer} when  HRF is applied to its different modules on the IEMOCAP, M$^3$ED and DAIC-WOZ datasets.}
  \label{tab:hrf_speechformer} 
  \begin{threeparttable}
      \begin{tabular}{cccccccccccc}
        \toprule
       \multirow{2}{*}{which modules use HRF [$\times$8]}    &\multicolumn{5}{c}{IEMOCAP}   & \multicolumn{3}{c}{M$^{3}$ED} & \multicolumn{3}{c}{DAIC-WOZ}\\
        \cmidrule(lr){2-6} \cmidrule(lr){7-9} \cmidrule(lr){10-12}
        &WA     &UA     &WF1    &Params &FLOPs   &WF1    &Params &FLOPs  &MF1   &Params &FLOPs \\
        \midrule
        Original     &.599 &.605 &.598 & 323K  &50M  &.383 & 874K  &135M &.720 & 327K  &100M\\
        Lightweight  &.532 &.533 &.528 &3K   &1M     &.358 & 3K  &1M    &.620 & 3K  &3M\\
        \midrule
        FFN1         &.537 &.538 &.533 &3K   &1M   &\textbf{.363} & 3K  &1M &.643 & 3K  &3M\\
        FFN2         &\textbf{.545} &\textbf{.547} &\textbf{.543} &3K   &1M    &.360 & 3K  &1M &.669 & 3K  &3M\\
        FFN1 \& FFN2  &.506 &.491 &.503 &3K   &1M   &.354 & 3K  &1M &\textbf{.735} & 3K  &3M\\
        QKV          &.534 &.532 &.531 &3K   &1M   &.361 & 3K  &1M &.709 & 3K  &3M\\
        Project      &.539 &.529 &.536 &3K   &1M   &.362 & 3K  &1M  &.720 & 3K  &3M\\
        CLS      &.530 &.525 &.527 &3K   &1M   &.361 & 3K  &1M  &.709 & 3K  &3M\\
        ALL          &.536 &.535 &.531 &3K   &1M   &.362 & 3K  &1M &.694 & 3K  &3M\\
      \bottomrule
    \end{tabular}
\end{threeparttable}
\end{table*}

\subsection{Baseline Models}
Three typical Transformer models in the speech domain were selected to examine the performance of the introduced Transformer re-parameterization, and are listed as follows. 
\begin{itemize}
\item \textbf{ConvTransformer}  \cite{yeh_transformer}: A classic Transformer architecture for speech recognition. On top of the classic Transformer, it employs a VGG-style convolutional network that is able to potentially extract high-level representations of speech signals and their relative positions. 
\item \textbf{Conformer} \cite{gulati_conformer}: A frequently used model in speech domain. The Conformer is a Macaron-style structure which exploits FFN before and after the self-attention module. 
\item \textbf{SpeechFormer} \cite{chen_speechformer}: A latest SOTA Transformer variant that extracts different levels (\ie frame, phoneme, and word) of acoustic representations for SER. Compared to the original SpeechFormer, we made slight modifications. That is, we insert a linear layer to map the high-dimensional features of model inputs into a smaller one, which makes the $d_{model}$ reduced. 
\end{itemize}

\subsection{Results and Discussions}\label{subsec:results}

As aforementioned in Section~\ref{subsec:overview}, we investigated the modules of FFN, QKV, Project, CLS, and ALL for both ConvTransformer and SpeechFormer, and the one additional module of FFN-M for Conformer. Again, since an FFN is composed of two FC layers, it can be divided into FFN1, FFN2, and FFN1 \& FFN2. It is the same with FFN-M (\ie FFN-M1, FFN-M2, and FFN-M1 \& FFN-M2).

Table~\ref{tab:hrf_convtransformer}, Table~\ref{tab:hrf_conformer}, and Table~\ref{tab:hrf_speechformer} present the results evaluated on the IEMOCAP, M$^{3}$ED and DAIC-WOZ datasets for ConvTransformer, Conformer, and SpeechFormer, respectively. All these results were obtained when the HRF expansion ratio is eight. First of all, we can see that the original Transformer models, some of which have already been compressed, have hundreds of thousands of parameters and FLOPs (\ie computational complexity). With such kinds of model parameters and FLOPs, it is challenging to deploy these models on IoT devices.  

For this reason, we largely compressed the original ConvTransformer, Conformer, and SpeechFormer into tiny ones by reducing the depth and width of these models (shown in the second row of Table~\ref{tab:hrf_convtransformer}-\ref{tab:hrf_speechformer}) to make it more feasible under the on-device scenario. However, we also observe that their performance is greatly degraded, which is consistent with our expectations.

When applying HRF to the three selected Transformers, it is found that HRF-plugged models outperform the ones without HRF layers in most cases.  Even in some cases, they are comparable to the original models. Taking ConvTransformer for example, the best WF1s of the HRF model on IEMOCAP and M$^3$ED datasets are 0.580 and 0.387, which are obviously higher than 0.550 and 0.372 for the lightweight model and are close to 0.586 and 0.399 for the original model, respectively. 
Similarly, on the DAIC-WOZ dataset, the best MF1 score of the HRF model is 0.580, markedly higher than 0.533 for the lightweight model and exceeds 0.572 for the original model.
Therefore, the Transformer re-parameterization process can generally well fill the performance gap between the large models and their lightweight versions. This is largely attributed to the expansion of model size which plays an important role in model performance as discussed in previous work~\cite{allen_learning,allen_convergence}. 

In addition, we also notice that SpeechFormer performs less well compared to the other two Transformers on IEMOCAP and M$^3$ED datasets. This is possibly due to the specific design of SpeechFormer with four stages (each stage has a different number of layers of Transformer blocks) to extract features at different levels of speech (\ie frames, phonemes, words, and utterances). However, the lightweight model has only one stage and can only extract features at the frame level, which leads to a decrease in the overall learning ability. 
On the contrary, SpeechFormer performs the best on the DAIC-WOZ dataset. This is because the DAIC-WOZ dataset has too few samples, and the parameter count of ConvTransformer and Conformer is approximately three times that of SpeechFormer, making it prone to overfitting issues. SpeechFormer, on the other hand, is easier to train.

Moreover, when comparing different modules of FFN, QKV, Project, CLS, and ALL with HRF, we can generally find that the HRF FNN performs better than the other modules in five out of six cases with ConvTransformer, Conformer, and SpeechFormer on IEMOCAP, M$^{3}$ED and DAIC-WOZ datasets. This finding basically suggests that FFN layers play a vital role in Transformer, which consecutively maps the representations from the low-level space to a high-level space in a non-linear way. However, it is worth noting that FFN often holds a large ratio of the total parameters of Transformers due to its full-connection characteristics. Therefore, it is important to balance the size of FFN and its overall performance, which further raises the importance of the introduced re-parameterization approach. 

We further observe that the HRF FFN2 is superior to the HRF FFN1 or HRF FFN1 \& FFN2 in most cases. This indicates that mapping representations from high dimension to low dimension often lose a lot of information. This lost information can be well captured by inserting an HRF layer. Besides, it can be seen that when inserting an HRF layer for QKV attention modules, less performance improvement is observed. This is possible because of the redundancy of classic QKV matrics, which is consistent with previous findings in this work~\cite{michel_sixteen,voita_analyzing}. Furthermore, we also notice that combining different modules for HRF does not seem to help much. One potential assumption might be that the inserted HRF layers have no activation functions, and thus they have limited capability for the model to learn non-linearly. Such a kind of linear transformation might not need several times in one Transformer block. 

In addition to the above experiments, we also investigated how well the selected Transformer models are to confirm that these models are reliable representatives for SER. We compared the original three models with some other recent studies on SER. It is important to note that our proposed models are designed for mobile/edge device scenarios, which prevents us to exploit large pre-trained models, such as Wav2vec and HuBERT, to extract high-level representations. 
To ensure fair comparisons between these models, we only considered handcrafted acoustic features, including Mel spectrum, Log-mel spectrum, LMFB, MFCC, etc. 
The results of these comparisons are shown in Table~\ref{tab:hrf_sota_iemocap}, Table~\ref{tab:hrf_sota_m3ed} and Table~\ref{tab:hrf_sota_daic} for IEMOCAP, M$^{3}$ED and DAIC-WOZ datasets. 
As seen in Table ~\ref{tab:hrf_sota_iemocap} and Table~\ref{tab:hrf_sota_daic}, when using manual features, the original models can achieve comparable performance to other work for IEMOCAP and DAIC-WOZ. However, for M$^{3}$ED, our baseline was not as good as the SOTA model~\cite{zhao_m3ed}. This is due to the pre-trained model utilized for feature extraction in~\cite{zhao_m3ed}. For a fair performance comparison, we further extracted speech features using HuBERT, and the results of all three selected baseline models were better than the models in~\cite{zhao_m3ed}'s. All these results indicate the reliability of the selected baselines.

\begin{table}[!t]
  \centering
  \caption{Performance comparison between our exemplified Transformers and other SOTA models on the \textit{IEMOCAP} dataset when using classic acoustic features.}
  \label{tab:hrf_sota_iemocap}
  \begin{threeparttable}
      \begin{tabular}{cccccccc}
        \toprule
       Methods    &Features   &WA     &UA     &WF1  \\
        \midrule
        SpeechFormer \cite{chen_speechformer} &Log-mel      &.582 &.598 &-  \\
        CNN-ELM-attention \cite{guo_representation}  &Spectrogram &.613 &.604 &- \\
        SAMS\cite{hou_semantic}  &LMFB &.615 &.596 &-    \\
        BLSTM+attention\cite{guo_emotion}  & IS10 &.560 &.564 &-    \\
        our ConvTansformer  &LMFB  &.589 &.602 &.585     \\
        our Conformer  &LMFB  &.573 &.572 &.571     \\
        our SpeechFormer  &LMFB  &.599 &.605 &.598     \\
      \bottomrule
    \end{tabular}
\end{threeparttable}
\end{table}

\begin{table}[!t]
  \centering
  \caption{Performance comparison between our exemplified Transformers and other SOTA models on the \textit{M$^3$ED} dataset when using high-level representations extracted from pre-trained models or classic acoustic features.}
  \label{tab:hrf_sota_m3ed}
  \begin{threeparttable}
      \begin{tabular}{ccc}
        \toprule
        Methods   &Features   &WF1   \\
        \midrule
        $M^{3}$ED \cite{zhao_m3ed}  &Wav2vec 2.0       &.480  \\
        our ConvTransformer  &HuBERT           &.506     \\
        our Conformer       &HuBERT           &.491     \\
        our SpeechFormer    &HuBERT           &.505     \\
        our ConvTransformer  &LMFB     &.383     \\
        our Conformer        &LMFB     &.378     \\
        our SpeechFormer     &LMFB     &.399     \\
      \bottomrule
    \end{tabular}
\end{threeparttable}
\end{table}

\begin{table}[!t]
  \centering
  \caption{Performance comparison between our exemplified Transformers and other SOTA models on the \textit{DAIC-WOZ} dataset when using classic acoustic features.}
  \label{tab:hrf_sota_daic}
  \begin{threeparttable}
      \begin{tabular}{ccc}
        \toprule
        Methods   &Features   &MF1   \\
        \midrule
        FVTC-CNN \cite{huang2020exploiting} &FVTC &.640  \\
        EmoAudioNet \cite{othmani2021towards} &MFCC+Spectrogram &.653  \\
        Hybrid CNN-SVM \cite{saidi2020hybrid} &Spectrogram  &.680  \\ 
        CNN-LSTM \cite{solieman2021detection}  &COVAREP       &.610  \\
        SpeechFormer \cite{chen_speechformer} &Log-mel &.627  \\
        our ConvTransformer  &LMFB     &.572     \\
        our Conformer        &LMFB     &.633     \\
        our SpeechFormer     &LMFB     &.720     \\
      \bottomrule
    \end{tabular}
\end{threeparttable}
\end{table}

\subsection{Ablation Studies}
\begin{figure*}[!th]
  \centering
  \includegraphics[width=\linewidth]{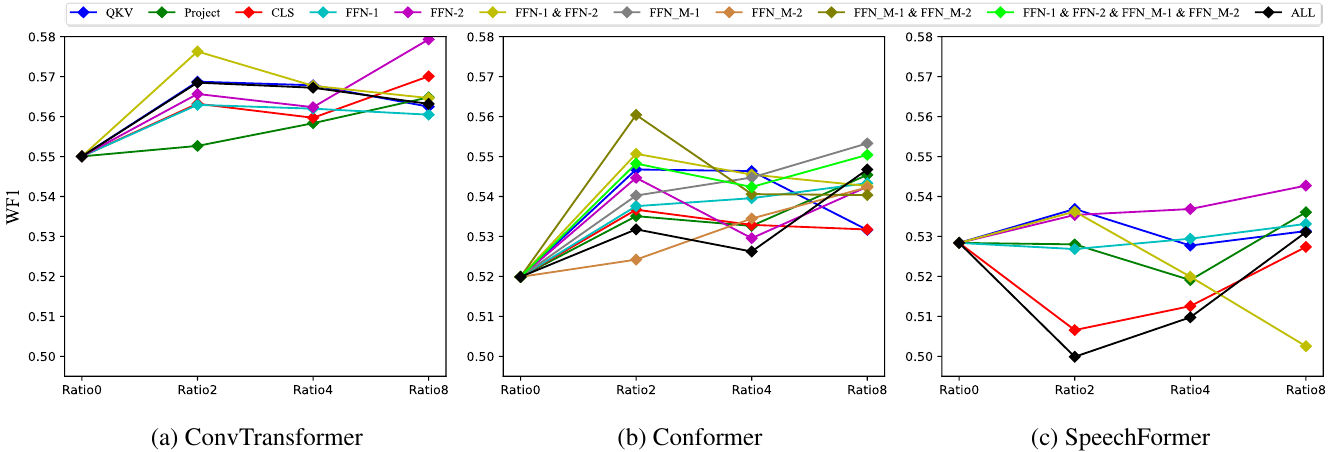}
  \caption{Results of WF1 when applying different \textit{expansion ratios} of HRF to different modules of ConvTransformer (a), Conformer (b), or SpeechFromer (c) on the \textit{IEMOCAP} dataset.}
  \label{fig:hrf_modules_iemocap}
\end{figure*}

\begin{figure*}[!th]
  \centering
  \includegraphics[width=\linewidth]{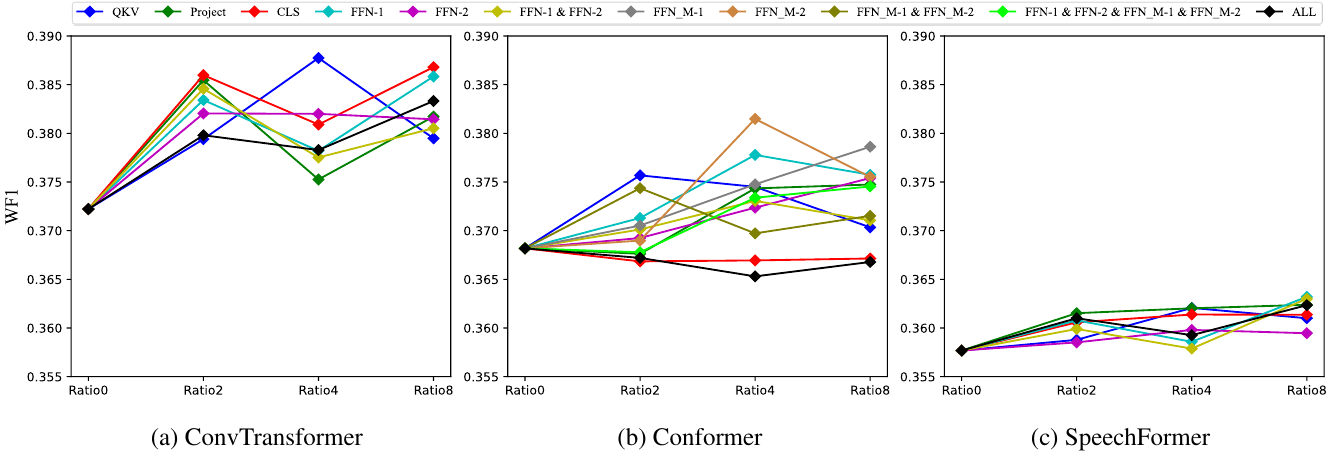}
  \caption{Results of WF1 when applying different \textit{expansion ratios} of HRF to different modules of ConvTransformer (a), Conformer (b), or SpeechFromer (c) on the \textit{M$^{3}$ED} dataset.}
  \label{fig:hrf_modules_m3ed}
\end{figure*}

\begin{figure*}[!th]
  \centering
  \includegraphics[width=\linewidth]{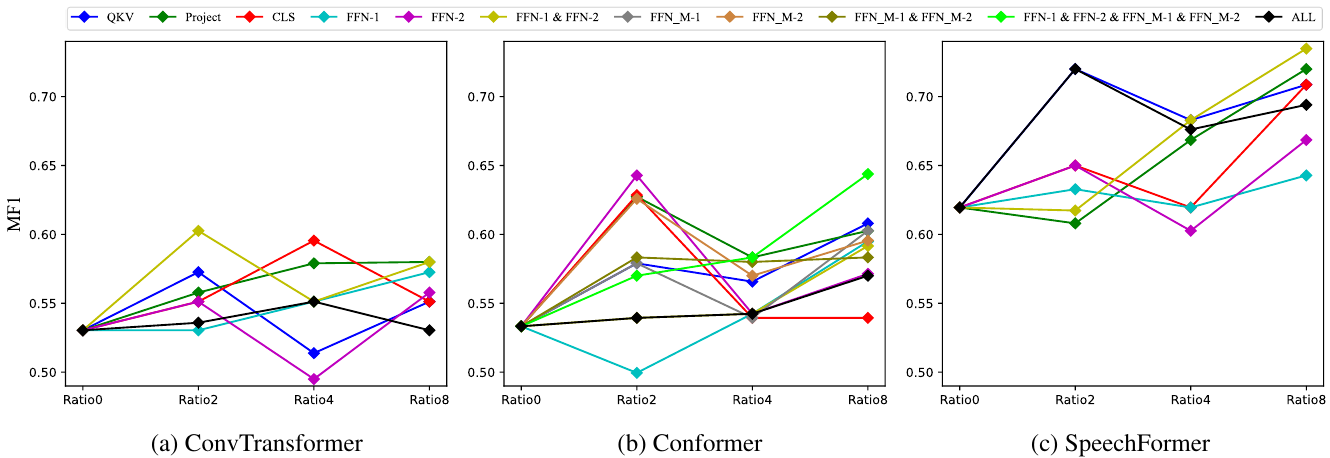}
  \caption{Results of MF1 when applying different \textit{expansion ratios} of HRF to different modules of ConvTransformer (a), Conformer (b), or SpeechFromer (c) on the \textit{DAIC-WOZ} dataset.}
  \label{fig:hrf_modules_daic}
\end{figure*}

\textbf{Impact of different expansion ratios ($r$) of HRF:} In order to examine the impact of HRF expansion at various ratios on the results, we conducted expansions of 0 (ratio0, original model), 2 (ratio2), 4 (ratio4), and 8 (ratio8) for all lightweight models in different modules. The obtained WF1 versus different expansion ratios of HRF for the three Transformers with different modules on the IEMOCAP, M$^{3}$ED and DAIC-WOZ datasets are plotted in Figure~\ref{fig:hrf_modules_iemocap}, Figure~\ref{fig:hrf_modules_m3ed} and Figure~\ref{fig:hrf_modules_daic} . 
When varying the expansion ratio, we can see that the HRF Transformers outperform their lightweight versions in most cases. This finding shows the robustness of the Transformer re-parameterization approach. Additionally, the different ratios display a slight double-increase trend, \ie when r=2, 4, or 8, the performance shows an increase, decrease, and increase. 
We hypothesize that this outcome indicates the phenomenon of double descent~\cite{d2020double, nakkiran_deep}, which has been proven to exist in image classification tasks. That is, the test error decreases, increases, and finally decreases again as the number of model parameters increases. 

 \begin{figure*}[!th] 
  \centering
  \includegraphics[width=\linewidth]{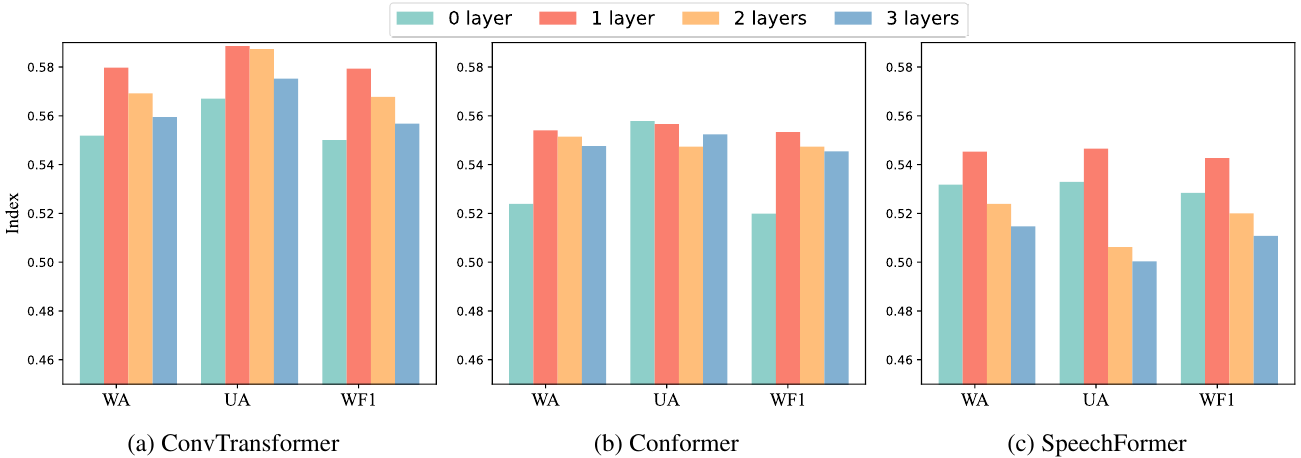}
  \caption{Results when applying different \textit{numbers of HRF layer} to the second feedforward layers of ConvTransformer (a), Conformer (b), or SpeechFromer (c) on the \textit{IEMOCAP} dataset.} 
  \label{fig:hrf_layers_iemocap}
\end{figure*}

\begin{figure*}[!th]
  \centering
  \includegraphics[width=\linewidth]{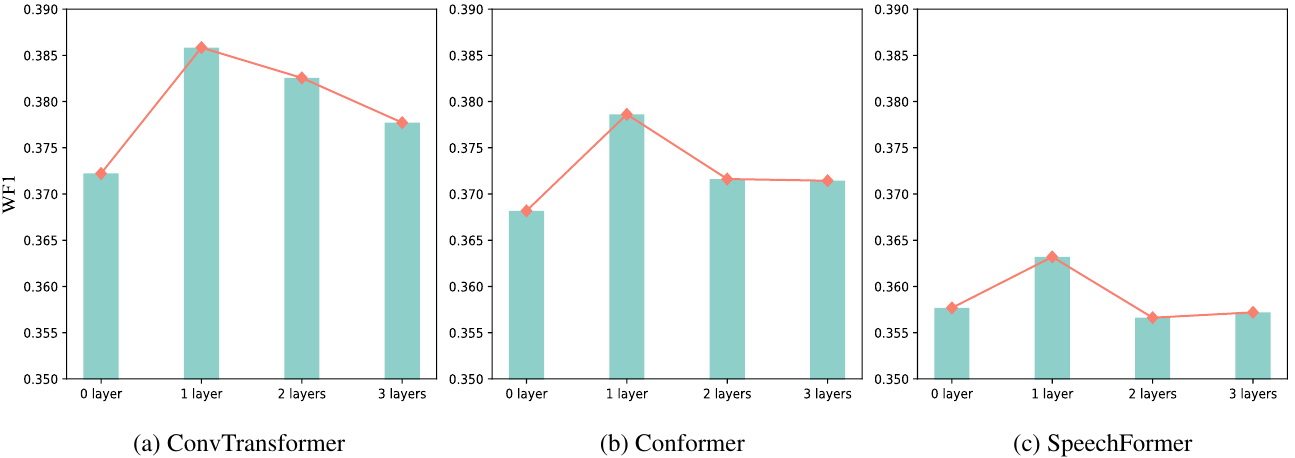}
  \caption{Results when applying different \textit{numbers of HRF layer} to the second feedforward layers of ConvTransformer (a), Conformer (b), or SpeechFromer (c) on the \textit{M$^{3}$ED} dataset.} 
  \label{fig:hrf_layers_m3ed}
\end{figure*}

\begin{figure*}[!th]
  \centering
  \includegraphics[width=\linewidth]{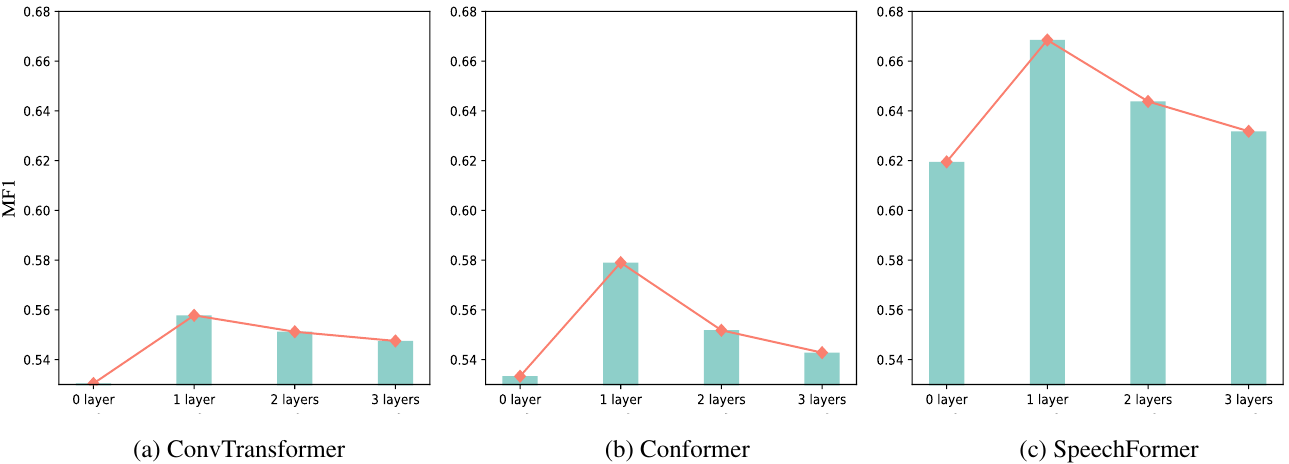}
  \caption{Results when applying different \textit{numbers of HRF layer} to the second feedforward layers of ConvTransformer (a), Conformer (b), or SpeechFromer (c) on the \textit{DAIC-WOZ} dataset.}
  \label{fig:hrf_layers_daic}
\end{figure*}

\textbf{Impact of different numbers of HRF layers ($n$):} 
To quantitatively analyze the impact of different numbers of HRF layers, we extend one HRF layer into two and three layers. To reduce the experiments, we took the best experimental setting for  ConvTransformer, Conformer, and SpeechFormer. That is, we fixed the HRF expansion ratio to eight in all cases. Then, for IEMOCAP, we applied multiple HRF layers to the modules of FFN2, FFN\_M1, and FFN2, for ConvTransformer, Conformer, and SpeechFormer, respectively; whilst for  M$^{3}$ED and DAIC-WOZ, we applied HRF to FFN1, FFN\_M1, and FFN1 for ConvTransformer, Conformer, and SpeechFormer, respectively. 
The results versus different numbers of HRF layers are illustrated in Figure~\ref{fig:hrf_layers_iemocap}, Figure~\ref{fig:hrf_layers_m3ed} and Figure~\ref{fig:hrf_layers_daic}. The results indicate that extending only one layer achieves the best results. This means that without a non-linear activation function, one HRF layer can be sufficiently well-trained. Increasing the depth of HRF layers may bring more noise and confusion into the system when doing gradient backpropagation. 

\begin{figure*}[!th]
  \centering
  \includegraphics[width=\linewidth]{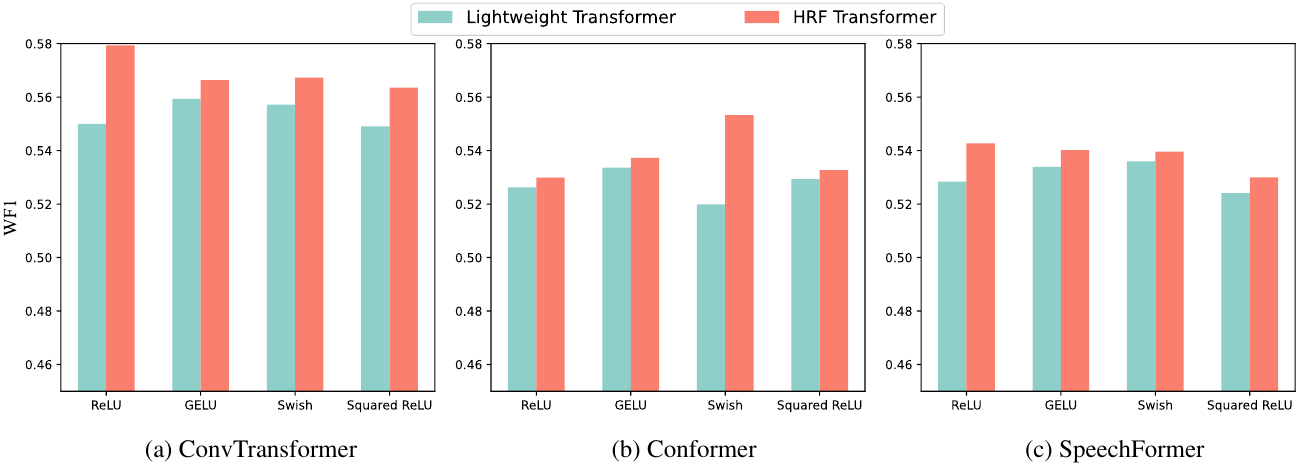}
  \caption{Performance of re-parameterized  lightweight ConvTransformer (a), Conformer (b), or SpeechFromer (c) with diverse \textit{activation functions} on the \textit{IEMOCAP} dataset.}
  \label{fig:hrf_activations_iemocap}
\end{figure*}

\begin{figure*}[!th]
  \centering
  \includegraphics[width=\linewidth]{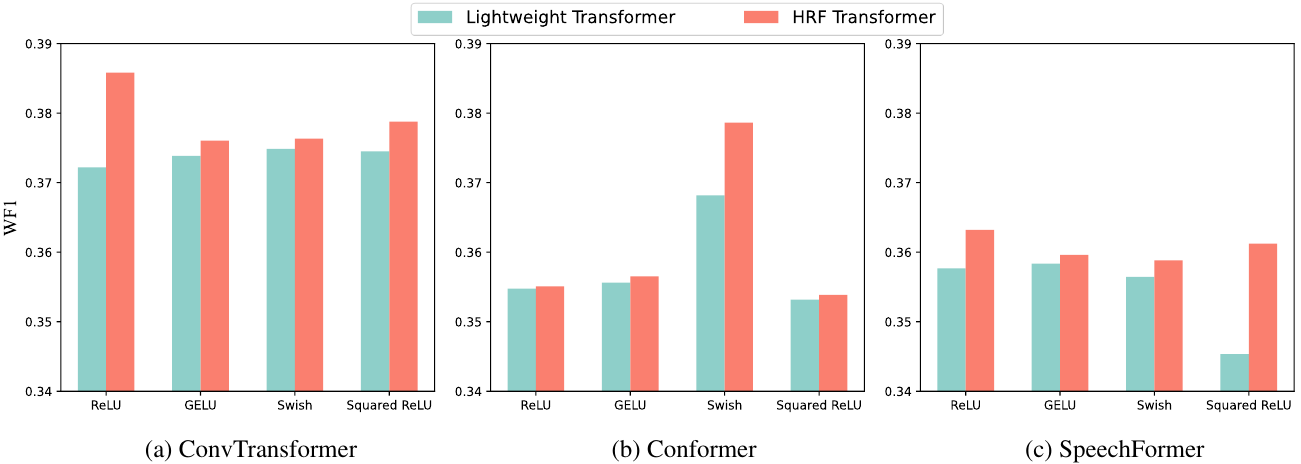}
  \caption{Performance of re-parameterized lightweight ConvTransformer (a), Conformer (b), or SpeechFromer (c) with diverse \textit{activation functions} on the \textit{M$^{3}$ED} dataset.}
  \label{fig:hrf_activations_m3ed}
\end{figure*}

\begin{figure*}[!th]
  \centering
  \includegraphics[width=\linewidth]{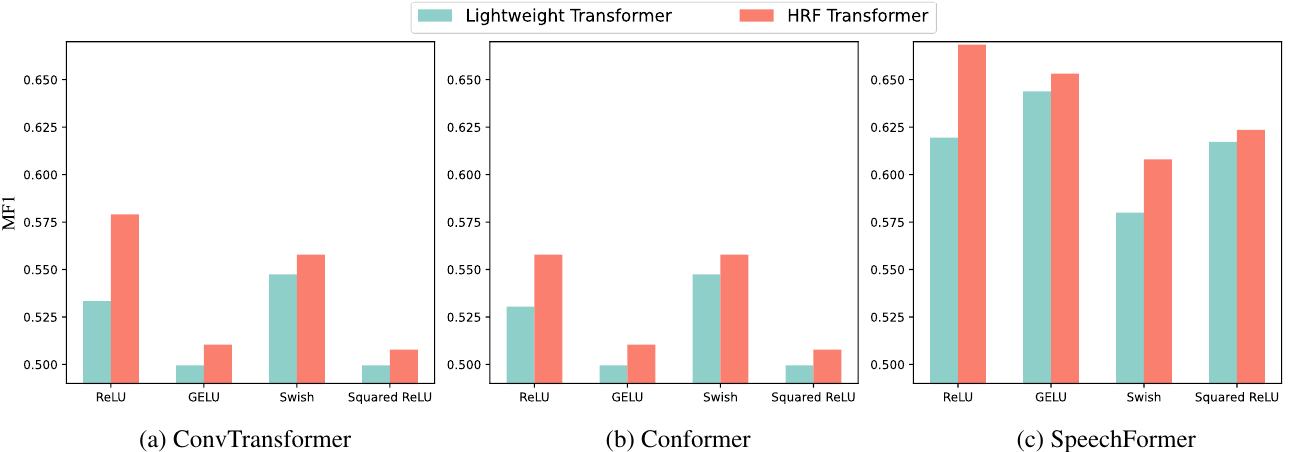}
  \caption{Performance of re-parameterized lightweight ConvTransformer (a), Conformer (b), or SpeechFromer (c) with diverse \textit{activation functions} on the \textit{DAIC-WOZ} dataset.}
  \label{fig:hrf_activations_daic}
\end{figure*}

\textbf{Impact of different activation functions for HRF:} Different activation functions were used for various Transformer variants. We tested four commonly used activation functions for the FFN layer, \ie ReLU, GELU, Swish, and Squared ReLU. The performance comparison between the lightweight Transformer and the HRF Transformer under different activation functions is depicted in Figure~\ref{fig:hrf_activations_iemocap}, Figure~\ref{fig:hrf_activations_m3ed} and Figure~\ref{fig:hrf_activations_daic}. 
It was observed that under all scenarios the HRF Transformer outperform the lightweight Transformer. 
This finding further demonstrates the robustness of the Transformer re-parameterization approach. 

\section{CONCLUSIONS}\label{section:5}
In the present paper, we introduce a structure re-parameterization method to boost the performance of compressed Transformers. The Transformer architecture can be expanded via a high-rank factorization process into a bigger model in the training, which helps increase the model learning capability.  In the inference, the expanded model is shrunk into the original one via a reverse mathematical calculation, while retaining the boosted model performance. Our approach is validated on the IEMOCAP, M$^{3}$ED and DAIC-WOZ datasets for the speech emotion recognition task.  Experimental results show the effectiveness of the structure re-parameterization method, which can enhance the lightweight Transformers, and even can make them comparable to larger models with significantly more parameters. The results also empirically reveal the robustness via investigating different numbers of expansion ratios and layers,  and different activation functions. 

Although the initial findings are promising, further investigation of the introduced method is necessary to unlock its full potential. Firstly, except for the speech modality in this work, other modalities of text or video will be explored as well for multimodal learning in future work. Secondly, we will extend the evaluation task into a broader range, such as image recognition, speech recognition, and target detection.



\ 

\ 

\

\

\bibliographystyle{IEEEtran}
\bibliography{IEEEabrv,dzr}

\vfill

\end{document}